\documentclass[final,5p,times,twocolumn]{elsarticle}

\usepackage{amssymb}
\usepackage{dsfont}
\usepackage{algorithm}
\usepackage{algorithmic}
\usepackage{xcolor}
\definecolor{revisioncolor}{RGB}{180,0,0}

\journal{Applied Energy}

\begin{document}

\begin{frontmatter}



\title{Microgrid Planner: A Distributed Energy Resource Sizing Method}


\author[inst1]{Daniel Reich}

\affiliation[inst1]{organization={Department of Operations Research, Naval Postgraduate School},%Department and Organization
            addressline={1 University Circle}, 
            city={Monterey},
            postcode={93955}, 
            state={California},
            country={USA}}

\begin{abstract}
We present a heuristic search method for distrubuted energy resource sizing, released in Microgrid Planner, an open-source software platform. Our method is constructed to identify a wide range of microgrid design options that satisfy a given set of power load requirements, allowing a decision maker to weigh trade-offs between potential designs and select preferred solutions. We introduce a global binary search algorithm to build a diverse set of microgrid design options and refine them using a local linear search method.
\end{abstract}

% \begin{highlights}
% \end{highlights}


\end{frontmatter}


\section{Introduction}
Microgrids are controllable systems for distributing power within a small geographic area, which could be a community, a company base or a military installation. A microgrid is composed of distributed energy resources (DERs), examples of which are diesel generators, photovoltaic systems, wind turbines and battery energy storage systems (BESS). A microgrid may be the sole energy source for an off-grid location; it may supplement the electrical grid; or it may be a backup in the event of a grid outage. Microgrid Planner \citep{reich2024platform} is an open-source software platform designed to deploy analytical methods for microgrid planning. In this paper, we present a DER sizing method released in Microgrid Planner.

The design of general-purpose microgrids has been an active area of 
research in recent years \citep{al2017review, khatib2016review}. A central element of designing a microgrid is selecting appropriate DER capacities. Particular emphasis has been placed on optimization-focused methods that provide a decision-maker with a single best solution for selecting the sizes of DERs \citep{zolan2021decomposing, alsaidan2017comprehensive, fossati2015method, 
bahmani2014optimal, chatterji2020battery, chen2011sizing, liu2019computationally, dong2018battery, 
al2014techno, rodriguez2017siting, hussain2019heuristic, zhao2014optimal, cao2020hybrid, bukar2019optimal, lan2015optimal, lai2019sizing, mashayekh2017mixed, anderson2022north}.

In the context of microgrids, objectives may include several factors, such as cost, sustainability and reliability. Combining these into a singular weighting that may then be optimized inherently results in a loss of information. Decision makers may prefer to review a range of options, weigh tradeoffs between factors they understand well, and then further investigate a small set of alternatives they deem worthy of consideration.

A decision maker intuitively may know that a range of potential solutions exists. For example, a decision maker may understand that at one end of the spectrum it is possible to produce all power required using diesel generators and at the other end of the spectrum it is possible to produce the same power using wind turbines and batteries. We prefer to empower the decision maker to play an active role in the decision process, by producing a wide range of microgrid designs with varying, appropriately sized combinations of DERs.

The method for DER sizing that we present in this paper is a more generalized heuristic for the ``rightsizing'' approach of \cite{reich2021rightsizing}. Their work identified all potential microgrid designs that could meet an input power load profile requirement. However, their search algorithm was limited to three dimensions, where they considered diesel generation, photovoltaic and battery energy storage systems. Microgrid Planner is intentionally configured to support any number of defined technologies, so it requires a more flexible sizing algorithm that can accommodate the entire set of options. Unlike the work of \cite{reich2021rightsizing}, our approach does not identify every potentially ``rightsized'' microgrid design, but rather a sufficiently varied subset of those designs.  

The remainder of this paper is organized as follows. In Section \ref{sec:metrics}, we formally define the problem and introduce performance metrics for candidate microgrid designs. In Section \ref{sec:our-sizing}, we present our DER sizing method. In Section \ref{sec:experiment}, we provide a computational experiment and results. We conclude with closing remarks in Section \ref{sec:conclusions}.

\section{Problem Statement and Metrics}\label{sec:metrics}
Our aim is to identify a diverse set of microgrid design options composed of DERs with sufficient capacities to meet power demand over a given time horizon. To state this formally, let us consider a function $f(g,p) = y \in \{0,1\}^{\vert T \vert}$ that maps an input microgrid design $g$ and power load $p$ to binary outcomes $y$, where $y_t=1$ if and only if power load $p_t$ is not fully satisfied over time interval $t \in T$ with duration $d_t$. Then $\bar{y} = \sum_{t \in T} y_t d_t / \sum_{t \in T} d_t$ provides a performance measure representing the proportion of time a power load is not fully satisfied. The function $f$ can take the form of an optimization or simulation model that incorporates the control logic for an energy management system.

We can define the notion of dominance between two microgrids $g^1$ and $g^2$ with respect to a given $f$ and $p$. Let $c^1_i$ and $c^2_i$ be the capacities of DER $i$ for microgrids $g^1$ and $g^2$, respectively. Then, $g^1$ dominates $g^2$, denoted $g^1 \succ g^2,$ if $\bar{y}^1 \leq \bar{y}^2$ and $c^1_i \leq c^2_i$ for every $i \in I$. In other words, if one microgrid performs at least as well as another without having more DER capacity, then it is dominant. The set of all ``rightsized'' microgrid designs can be stated as $S=\{g^1, \cdots g^m\}$ such that $g^j \nsucc g^k$ for any $(g^j, g^k) \in S \times S$, $j\neq k$. $S$ is a finite set because capacities defining DERs are discrete. Our goal is to find a diverse subset of $S$, where diversity can be measured by the difference between capacities of any two microgrids $g^j$ and $g^k$. 

\begin{figure*}
    \centering
    \includegraphics[width=6.5in]{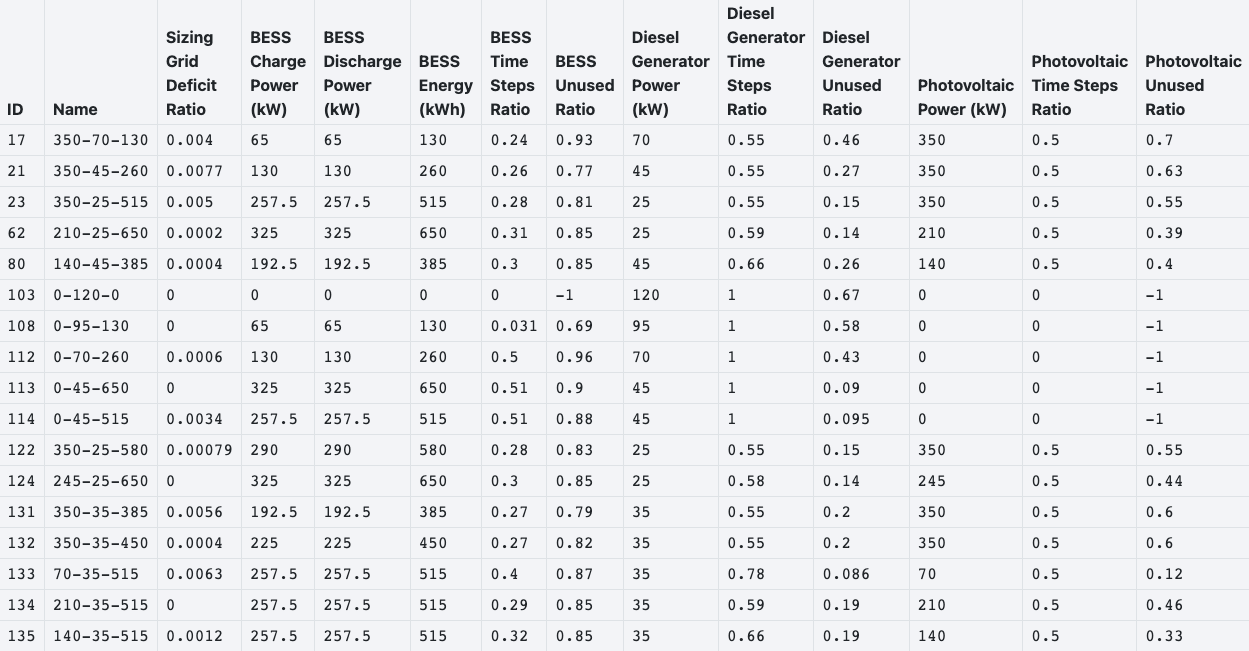}
    \caption{An example of non-dominated microgrid designs generated by our DER sizing method. Ratio values equal to -1 correspond to DER capacities of 0.}
    \label{fig:sizing-results}
\end{figure*}

Figure \ref{fig:sizing-results} shows the results of the DER sizing algorithm in the Microgrid Planner web app. The column ``Sizing Grid Deficit Ratio'' provides the performance measure $\bar{y}$. Practical candidate solutions have a value of 0 or very close to 0, so we filter out solutions over a selected value, which in the example shown is set to 0.01. While this value may be too high for critical applications, it may be suitable in some contexts, so we use it for illustrative purposes. Dominated solutions are not shown to the user.

Detecting whether excess capacity exists for a potential microgrid design is not as straightforward as detecting deficits, because excess power is to be expected at many points in time in order to have sufficient capacity at the times with highest power demand or lowest power availability. We compute an individual performance metric for each component in any candidate microgrid design we identify, shown in Figure \ref{fig:sizing-results} in the columns ``\{DER NAME\} Unused Ratio''. This ratio measures the proportion of time steps where power from a given DER was available but not used. The metric is nuanced because not all DER types may supply power at all times, for example, photovoltaic systems are useful only during daylight hours; so to measure the utilization, we limit the domain of consideration to those daylight hours and calculate the unused proportion accordingly. %A second complicating factor for this metric is its dependence on energy management system logic, which controls the prioritization of power sources at all points in time. Energy management systems are both decoupled and configurable within Microgrid Planner and are not the focus of this paper, so we refer the reader to \cite{reich2021rightsizing} for additional detail. Herein, we treat them abstractly as part of our function $f$.

\section{Sizing Method}\label{sec:our-sizing}
%a microgrid, which defines values for all DER component attributes, for example battery minimum state of charge. These attribute values are retained throughout the method with the exception of capacity levels that are modified during the search procedures. Other inputs are 
Our microgrid sizing method takes as input a function that operates the microgrid, a power load profile, DER capacity bounds, and both a small and larger number of discrete capacity levels to consider. Our sizing method can be broken down into a three-step process:
\begin{enumerate}
    \item Generate an initial set of microgrids, by exhaustive search using a small number of capacity levels.
    \item Generate a more diverse set of microgrids, by executing a binary search algorithm using a larger number of capacity levels, and using the initial set of microgrids identified in prior step.
    \item Generate a refined set of microgrids by executing a local search algorithm for each non-dominated microgrid identified in the prior step, skipping ones with power deficits.
\end{enumerate}

\begin{algorithm}
\caption{\label{alg:exhaustive-search}Microgrid sizing exhaustive search algorithm.}
\begin{algorithmic}\fontsize{10}{11.2}\selectfont
\renewcommand{\algorithmicrequire}{\textbf{Input:}}
\renewcommand{\algorithmicensure}{\textbf{Output:}}
\REQUIRE $f$ (function for operating microgrid)
\REQUIRE $p$ (power load, indexed by time)
\REQUIRE $l_i, i \in I$ (capacity lower bounds by DER)
\REQUIRE $u_i, i \in I$ (capacity upper bounds by DER)
\REQUIRE $n_i, i \in I$ (number of capacity levels by DER) 
\ENSURE  $S^1 = \{g^1, \cdots, g^m\}$ (set of microgrids)
\FOR{$i \gets 1$ to $\vert I \vert$}
    \STATE $c_i = \frac{u_i - l_i}{n_i}$
    \STATE $s^i = \{l_i + n_ic_i, \cdots, l_i + 2c_i, l_i+c_i, l_i\}$ 
\ENDFOR
\STATE $S^0 = s^1 \times s^2 \times s^{\vert I \vert}$
\STATE $S^1 = \{\}$
\STATE $S^2 = \{\}$
\FOR{$g \in S^0$}
    \FOR{$i \gets 1$ to $\vert I \vert$}
        \STATE $g' = g$
        \STATE set $c'_i$ to capacity of DER $i$ in $g'$
        \STATE $c''_i=\min\{c'_i+c_i,u_i\}$
        \STATE set capacity of DER $i$ in $g'$ to $c''_i$ to obtain $g''$
        \IF{$g'' \in S^2$}
            \STATE $S^2 = S^2 \cup \{g\}$
            \STATE break
        \ENDIF
    \ENDFOR
    \IF{$g \in S^2$}
        \STATE continue
    \ENDIF
    \STATE $y^{g} = f(g,p)$
    \STATE $S^1 = S^1 \cup \{g\}$
    \IF{$\bar{y}^{g}>0$}
        \STATE $S^2 = S^2 \cup \{g\}$
    \ENDIF
\ENDFOR
\RETURN $S^1$
\end{algorithmic}
\end{algorithm}

Our exhaustive search method is presented in Algorithm \ref{alg:exhaustive-search}. Potential microgrid designs are not explored further when ones with equal or greater capacities for all DERs have already been found insufficient to meet power demands at all points in time.\footnote{Counterintuitively, this could result in pruning a solution that actually meets power demand, due to subtleties in the battery charge states resulting from the energy management system logic. However, such cases are rare.}

\begin{algorithm}
\caption{\label{alg:sizing-binary-search}Microgrid sizing binary search algorithm.}
\begin{algorithmic}\fontsize{10}{11.2}\selectfont
\renewcommand{\algorithmicrequire}{\textbf{Input:}}
\renewcommand{\algorithmicensure}{\textbf{Output:}}
\REQUIRE $f$ (function for operating microgrid)
\REQUIRE $p$ (power load, indexed by time)
\REQUIRE $l_i, i \in I$ (capacity lower bounds by DER)
\REQUIRE $u_i, i \in I$ (capacity upper bounds by DER)
\REQUIRE $n_i, i \in I$ (number of capacity levels by DER) 
\REQUIRE $S^0 = \{g^1, \cdots, g^m\}$  (set of microgrids)
\ENSURE  $S^1 = \{g^1, \cdots, g^{m'}\}$ (set of microgrids)
\STATE $S^1 = S^0$
\FOR{$g \in S^0$}
    \FOR{$i \gets 1$ to $\vert I \vert$}
        \STATE $c_i = \frac{u_i - l_i}{n_i}$
        \STATE $s^i = \{l_i, l_i+c_i, l_i + 2c_i, \cdots, l_i + n_ic_i \}$ 
        \STATE update capacity of DER $i$ in $g$ to closest capacity in $s^i$
    \ENDFOR
    \STATE $y^g = f(g,p)$
    \FOR{$l \gets 1$ to $\vert I \vert$}
        \IF{$\bar{y}^g==0$}
            \STATE decrease\_capacity = \TRUE
        \ELSE
            \STATE decrease\_capacity = \FALSE
        \ENDIF
        \STATE $g' = g$
        \FOR{$i \gets 1$ to $\vert I \vert$ in a random order}
            \STATE $h = 2^{\lfloor \log(n_i) \rfloor}$
            \WHILE{$h >= 1$}
                \STATE $g'' = g'$
                \WHILE{True}
                    \STATE $g' = g''$
                    \STATE set $c'_i$ to capacity of DER $i$ in $g'$
                    \IF{decrease\_capacity}
                        \STATE $c''_i=\max\{c'_i-hc_i,l_i\}$ 
                    \ELSE
                        \STATE $c''_i=\min\{c'_i+hc,u_i\}$ 
                    \ENDIF
                    \STATE set capacity of DER $i$ in $g'$ to $c''_i$ to obtain $g''$
                    \IF{$c''_i$ $!=$ $c'_i$}
                        \STATE $y^{g''} = f(g'',p)$
                        \STATE $S^1 = S^1 \cup g''$
                        \IF{$\bar{y}^{g''} > \bar{y}^{g'}$}
                            \STATE \textbf{break}
                        \ENDIF
                    \ELSE
                        \IF{\textbf{not} decrease\_capacity \AND $\bar{y}^{g'}==0$}
                            \STATE decrease\_capacity = \TRUE
                        \ENDIF
                        \STATE \textbf{break}
                    \ENDIF
                \ENDWHILE
                \STATE $h = h/2$
            \ENDWHILE
        \ENDFOR
    \ENDFOR
\ENDFOR
\RETURN $S^1$
\end{algorithmic}
\end{algorithm}

Our binary search procedure is presented in Algorithm \ref{alg:sizing-binary-search}. If a power deficit is detected, then capacity is increased until no deficit exists, at which point the search direction is reversed; otherwise, capacity is decreased. A capacity adjustment is made on only one DER component per iteration. The step size is initialized to the power of 2 less than the number of capacity levels; it is halved whenever a deficit is encountered until the step size reaches 1, the stopping condition. The second and third level outer for-loops allow the binary searches to be run for DERs in differening orders to explore multiple branches of the exhaustive search tree.

\begin{algorithm}
\caption{\label{alg:sizing-local-search}Microgrid sizing local search algorithm.}
\begin{algorithmic}\fontsize{10}{11.2}\selectfont
\renewcommand{\algorithmicrequire}{\textbf{Input:}}
\renewcommand{\algorithmicensure}{\textbf{Output:}}
\REQUIRE $f$ (function for operating microgrid)
\REQUIRE $p$ (power load, indexed by time)
\REQUIRE $l_i, i \in I$ (capacity lower bounds by DER)
\REQUIRE $u_i, i \in I$ (capacity upper bounds by DER)
\REQUIRE $n_i, i \in I$ (number of capacity levels by DER) 
\REQUIRE $S^0 = \{g^1, \cdots, g^m\}$  (set of non-dominated microgrids)
\ENSURE  $S^1 = \{g^1, \cdots, g^{m'}\}$ (set of microgrids)
\STATE $S^1 = S^0$
\FOR{$g \in S^0$}
    \STATE $y^g = f(g,p)$
    \IF{$\bar{y}^g>0$}
        \STATE \textbf{continue}
    \ENDIF
    \STATE $g' = g$
    \FOR{$l \gets 1$ to $\vert I \vert$}
        \FOR{$i \gets 1$ to $\vert I \vert$}
            \STATE $c_i = \frac{u_i - l_i}{n_i}$
            \WHILE{True}
                \STATE set $c'_i$ to capacity of DER $i$ in $g'$
                \STATE $c''_i=\max\{c'_i-c_i,l_i\}$ 
                \IF{$c''_i==c'_i$}
                    \STATE \textbf{break}
                \ENDIF
                \STATE set capacity of DER $i$ in $g'$ to $c_i''$ to obtain $g''$
                \STATE $y^{g''} = f(g'',p)$
                \STATE $S^1 = S^1 \cup g''$
                \IF{$\bar{y}^{g''} > 0$ }
                    \STATE \textbf{break}
                \ENDIF
                \STATE $g' = g''$
            \ENDWHILE
        \ENDFOR
    \ENDFOR
\ENDFOR
\RETURN $S^1$
\end{algorithmic}
\end{algorithm}

In the final phase of our heuristic, all the non-dominated solutions generated by Algorithm \ref{alg:sizing-binary-search}, aside from those with power deficits, are finessed through our local search procedure presented in Algorithm \ref{alg:sizing-local-search}. Capacity is decreased until either no further decrease is possible or until a power deficit arises. A capacity adjustment is made on only one DER component per iteration. The second and third level outer for-loops allow the local searches to be run for DERs consecutively until the stopping condition is reached. 

\section{Computational Experiment}\label{sec:experiment}

Microgrid Planner is intentionally designed for researchers to update DER models with ease, so the specific assumptions regarding the operation of DERs that are employed in our computational experiment are not central to the main focus of this paper. We therefore refer the reader to \cite{reich2021rightsizing} for details on the simulation method and DER models we employ from Microgrid Planner\footnote{We use version 1.1, prior to the integration of historical weather data in version 2.0.}

\begin{figure*}
    \centering
    \includegraphics[width=6.5in]{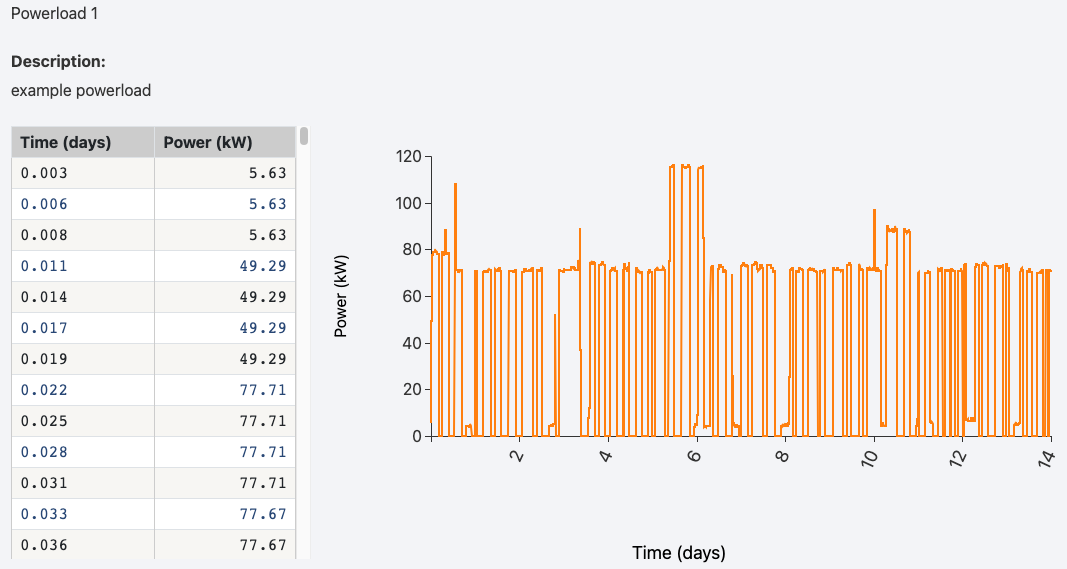}
    \caption{An example power load profile with 5040 time steps covering a 2-week horizon in 4-minute intervals.}
    \label{fig:powerload}
\end{figure*}

Our computational example is motivated by a military deployment with planned duration of two weeks, using the baseline power load shown in Figure \ref{fig:powerload}, which is the same as in \cite{reich2021rightsizing} and \cite{reich2023sensitivity}. The time horizon is discretized into 4-minute intervals, which translates to 5040 time steps at which microgrid operation is simulated. If substituting hourly data that is a typical discretization in commercial software products, the same problem size would accommodate a 7-month planning horizon.

We set lower bounds of 0 for all DER capacities.\footnote{Our method treats BESS energy as the primary BESS capacity and maintains energy-to-charge-power and energy-to-discharge-power ratios from an input microgrid template in Microgrid Planner.}  We scale the upper bounds by the peak power demand in the power load profile with adjustments to account for losses. For diesel generators and wind turbine, we apply a multiplier of 1. For photovoltaic systems, we apply a multiplier of 3 to compensate for non-peak and night hours with less or no generation. For battery energy storage systems, we apply a multiplier of 5 to allow for high levels of energy reserves. We test parameter settings of 11, 21, 41, 81 and 161 levels with respective capacity spacing of 10\%, 5\%, 2.5\%, 1.25\% and 0.625\% of the allowable maximum for each DER type. For generating the initial solution set from the exhaustive search algorithm, we use 6 levels with respective capacity spacing of 20\%. We also attempt to solve the exhaustive search algorithm with more levels, for comparison purposes, but it becomes computationally intractable.

\subsection{Computational Results}

\begin{figure*}
    \centering
    \includegraphics[width=6.5in]{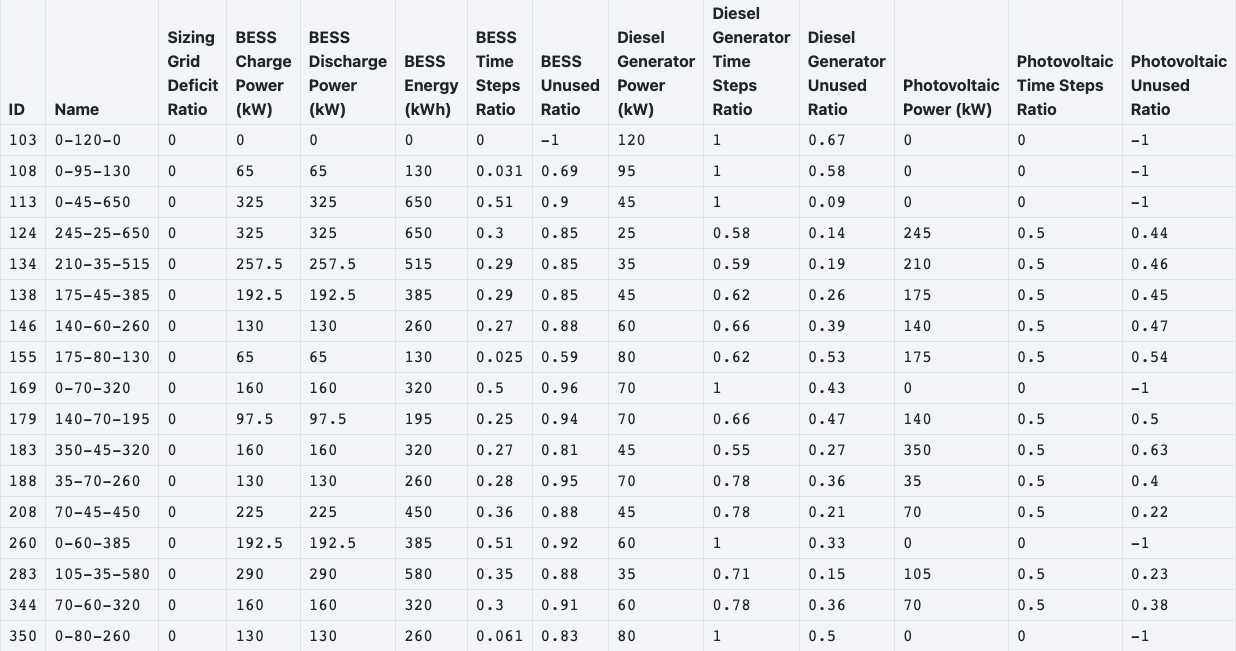}
    \caption{An example of non-dominated microgrid designs with no power deficit generated by our DER sizing method. Ratio values equal to -1 correspond to DER capacities of 0.}
    \label{fig:sizing-results-no-deficit}
\end{figure*}

Our heuristic sizing algorithm with 11 levels (10\% spacing) simulates 359 candidate microgrid designs when considering three DER types: diesel generators, photovoltaic and battery energy storage systems. Of these, the 17 solutions shown in Figure \ref{fig:sizing-results-no-deficit} are all the non-dominated solutions with no power deficits at any point in time. The diversity of these candidate solutions is evident in the number of discrete capacity levels they cover. Specifically, 8 out of 11 diesel generation capacities, 9 out of 11 photovoltaic system capacities, and 10 out of 11 battery energy storage system capacities are present in these 17 candidate microgrid designs. We are able to compare the results from our sizing method with those obtained from an exhaustive search because the instance is sufficiently small. Exhaustive search produces 18 solutions with no power deficits and not identified as dominated, 16 of which are also identified by the heuristic. The heuristic misses two solutions that the exhaustive search identifies, one of which exhaustive search prunes, because of a subtle detail regarding energy management system logic and resulting battery state of charge. 

\begin{table*}
    \centering
    \begin{tabular}{c|c|c||c|c|c||c|c|c}
        \multicolumn{3}{c||}{Sizing Method} & \multicolumn{3}{c||}{Reich \& Oriti \cite{reich2021rightsizing}} & \multicolumn{3}{c}{Difference} \\
        \hline
         diesel & photovoltaic & battery & diesel & photovoltaic & battery & diesel & photovoltaic & battery \\
         \hline
         25  & 245 & 650 & 25 & 240 & 635 & 0 & 5 & 15 \\
         35  & 105 & 580 & 35 & 100 & 570 & 0 & 5 & 10 \\
         35  & 210 & 515 & 35 & 210 & 500 & 0 & 0 & 15 \\
         45  & 0 & 650 &  45 & 0 & 610 & 0 & 0 & 40 \\         
         45  & 70 & 450 &  45 & 65 & 455 & 0 & 5 & -5 \\
         45  & 175 & 385 & 45 & 170 & 370 & 0 & 5 & 15 \\
         45  & 350 & 320 & 45 & 335 & 320 & 0 & 15 & 0 \\
         60  & 0 & 385 & 55 & 0 & 360 & 5 & 0 & 25 \\
         60  & 70 & 320 & 55 & 70 & 315& 5 & 0 & 5 \\
         60 & 140 & 260 & 60 & 140 & 240 & 0 & 0 & 20 \\
         70  & 0 & 320 & 70 & 0 & 275 & 0 & 0 & 45 \\
         70 & 35 & 260 & 70 & 35 & 250 & 0 & 0 & 10 \\
         70  & 140 & 195 & 70 & 140 & 185 & 0 & 0 & 10 \\
         80  & 0 & 260 &  75 & 0 & 245 & 5 & 0 & 15 \\
         80  & 175 & 130 & 80 & 160 & 130 & 0 & 15 & 0 \\
         95  & 0 & 130 & 95 & 0 & 125 & 0 & 0 & 5 \\
         120  & 0 & 0 & 120 & 0 & 0 & 0 & 0 & 0 \\
    \end{tabular}
    \caption{Comparison of candidate microgrid designs produced by our sizing method versus their closest counterparts using the ``rightsizing'' method of \cite{reich2021rightsizing}.}
    \label{tab:sizing-results-comparison}
\end{table*}

We compare our sizing algorithm results with those of the nested binary search ``rightsizing'' method presented in  \cite{reich2021rightsizing} with step sizes of 5 for all DER type capacities. The results are shown in Table \ref{tab:sizing-results-comparison}. The difference between solutions generated with the two approaches is less than the 10\% spacing set by the input number of levels, 11. For example, the largest difference encountered is in the microgrid with 70 kW diesel generation and no photovoltaic system, where our sizing method sized the battery energy storage system at 320 kWh, which is 45 kWh more than the method of \cite{reich2021rightsizing}. This difference may seem large at first glance. However, when considered in context of the discretization of battery capacity levels in our sizing method, it actually represents the lowest possible level corresponding to the ``rightsized'' value, because decreasing another level to 260 kWh results in a non-zero power deficit. The same is true for all differences reported. In other words, given the discretization of capacity levels we used in our computations, the 17 non-dominated microgrids are all ``rightsized''.

\begin{table*}
    \centering
    \begin{tabular}{c|c|c||c|c|c||c|c|c}
        \multicolumn{3}{c||}{} & \multicolumn{3}{c||}{Sizing Method} & \multicolumn{3}{c}{Exhaustive Search} \\
        \hline
         \# DER & \# levels & \# potential & \# candidate & \# sim & run & \# candidate & \# sim & run \\
         types & per DER & microgrid & microgrid & & time & microgrid & & time \\
          &  & designs & designs & & (min) & designs & & (mins)  \\
         \hline
         3  & 6 & 216 & N/A & N/A & N/A & 7 & 114 & 2\\
         3  & 11 & 1331 & 17 & 359 & 8 & 18 & 665 & 14 \\
         3  & 21 & 9261 & 33 & 615 & 15 & 62 & 4562 & 101 \\
         3  & 41 & 68,921 & 42 & 816 & 20 & 132 & 21,324 & 770\\
         3  & 81 & 531,441 & 42 & 920 & 23 & - & - & - \\
         3  & 161 & 4,173,281 & 44 & 1160 & 28 & - & - & -\\ 
         4  & 6 & 1296 & N/A & N/A & N/A & 29 & 1104 & 33 \\
         4  & 11 & 14,641 & 90 & 2196 & 64 & 132 & 12,691 & 295 \\
         4  & 21 & 194,481 & 176 & 3711 & 109 & - & - & -\\
         4  & 41 & 2,825,761 & 223 & 5652 & 168 & - & - & - \\
         4  & 81 & 43,046,721 & 236 & 6480 & 188 & - & - & - \\
         4  & 161 & 671,898,241 & 254 & 9287 & 216 & - & - & -\\
    \end{tabular}
    \caption{Summary of our sizing method versus exhaustive search for varying numbers of DER types and levels per DER type. ``N/A'' denotes 6-level per DER type instances for which our sizing method is not applicable, because it would return the exhaustive search results. ``-'' denotes instances for which exhaustive search was not solvable within the 48-hour time limit.}
    \label{tab:sizing-results-by-levels}
\end{table*}

To further reduce the differences between the solutions generated by the two methods, we must reduce the capacity level spacing by increasing the number of levels per DER. Reducing the capacity level spacing also increases the number of potential ``rightsized'' solutions. A summary of results for the various capacity level spacings we tested is provided in Table \ref{tab:sizing-results-by-levels}. 42 out of the 44 solutions obtained by our sizing method when using 161 levels per DER match exactly to solutions generated by the the method of \cite{reich2021rightsizing}. The other two differ by only 5 kWh of battery capacity, the rounding precision set for generating the discrete capacity levels to be considered. A similar comparison is not possible in the four DER type instance, because the the method of \cite{reich2021rightsizing} is limited to three DER dimensions.

Exhaustive search exceeded the 48-hour time limit with 81 levels for three DER types and with only 21 levels with four DER types. The run time of our sizing method increases at a decreasing rate as the number of levels are increased at an exponential rate, so it scales well to achieve capacity precision. However, it does not scale as well when increasing the number of DER types. Microgrid Planner does not yet include a fifth type of DER, so future work may be required to improve the efficiency when more DER types are supported. Avenues exist for doing so, such as decreasing the number of levels in the exhaustive search used to provide the initial set of microgrid designs, reducing the number of iterations in the for-loops currently used to expand the search and pruning more solutions.

\section{Conclusions}\label{sec:conclusions}
We developed the sizing method we introduce in this paper to meet the planning needs of our partners at military installations. We construct search heuristics that are efficient for the size of the current DER technology portfolios being actively planned for current microgrid investments. These heuristics are customizable and may be tuned as additional technologies are considered; however, we envision the DER technology set to remain relatively small in the near future. Our method has been released in Microgrid Planner \citep{reich2024platform}.

\section*{Acknowlegements}
This work was supported by the NextSTEP Program, sponsored by the Office of Naval Research; by Naval Facilities Engineering Systems Command (NAVFAC) as part of the Navy Shore Energy Technology Transition and Integration program; and by the Director of Operational Energy, Deputy Assistant Secretary of the U.S. Navy. The authors thank Leah Frye for contributions to the development of the frontend user interface. DoD Distribution Statement: Approved for public release; distribution is unlimited. The views expressed in this document are those of the 
authors and do not necessarily reflect the official policy or position of the DoD or the U.S. Government. 

\bibliographystyle{elsarticle-num} % outcomment this and next line in Case 1
\bibliography{microgrid.bib} % if more than one, comma separated





% \end{thebibliography}
\end{document}